\documentstyle[12pt]{article}
\begin{document}
\title{The Astronomical Link Betweeen India and the Mayans}
\author{B.G. Sidharth$^*$\\
B.M. Birla Science Centre, Adarsh Nagar, Hyderabad - 500 063 (India)}
\date{}
\maketitle
\footnotetext{$^*$Email:birlasc@hd1.vsnl.net.in; birlard@ap.nic.in}
\begin{abstract}
We argue that there was a link between Indus Valley India and the Mayans of Central
America which is brought out by astronomical references. The former used a 
Jovian calendar while the latter had perfected a
calendar based on Venus. This will be shown to be a significant clue. 
This also provides an explanation for the as yet unexplained
fact that the Indus script has been found on Easter Islands.
\end{abstract}
Over the past several decades, the Indian, more accurately Hindu influence
on the Mayan civilization of Central America has been noted by many scholars
\cite{r1}. It is unanimously agreed that there was direct or indirect
contact over the centuries, through South East Asia, China and the Far East,
though exactly when and how this took place is not as yet clear. As is well
known, the Mayan civilization which goes back to about 2600 B.C., or earlier,
had an advanced astronomy and calendar tradition\cite{r2}.\\
We mention some of the Hindu influences \cite{r1,r3}:\\
1.  Figures of the Hindu cosmic myth known popularly as the churning of the
ocean are found in the Mayan documents. The astronomical significance of the
myth has been explained in reference\cite{r4}. Infact a very early depiction of this myth is to be found in the
Angkor Vat temple in present day Cambodia.\\
2.  The Swastika symbol, interpreted by some as representing the cosmic cycle,
was also present there.\\
3.  Motifs of the lotus, elephant, sea goat and the like, unmistakably paint
to a Hindu influence.\\
4.  Unmistakable architectural similarities and parallels exist, though, 
according to Hindu references, in this case it might well be the Mayan influence
on Hindu architecture.\\
It is also interesting to note that the Mayan calendar began with a date
around 3112 B.C., very close to the Hindu traditional beginning of the Kali
age viz., 3102 B.C.\cite{r5,r6}. This period in India corresponds to the
Indus valley or Harappan civilization which lasted for a few thousand years,
atleast up to the first millennium B.C., or even later. The period around
the first millennium B.C. is also the period of the epic Mahabharata. Marine
archaeological researches in the Dwaraka region of Western India put this
period near 1500 B.C. \cite{r7}. Infact the Mahabharata period of India
overlapped the Indus valley period\cite{r8}. Arjuna one of the heroes of the
Mahabharata was a friend of Maya, an expert architect and he had also married
a Naga princess, two facts, which, as will be seen now could be of
significance. Maya himself is described as an Asura, as contrasted with
Devas (literally bright ones), an other fact of significance.\\
The vast body of what are called Puranas, a combination of obscure history
and mythology are clear on the point  that the Devas or the bright ones and
the Asuras, literally those who are deprived of the draught (of sun light),
are really anti podal inhabitants on the earth, or more precisely those,
whose longitude is separated by 180 degrees \cite{r9}. This is also explicitly
spelt out by Varaha Mihira\cite{r10}. Around 500, Varaha Mihira compiled some
of the then surviving ancient astronomical systems, and he mentions that when it is
noon for the Devas, it is mid night for the Asuras, and that they stand with
their heads down as compared to us, but that for them, we stand with our
heads down and so on.\\
Significantly, according to the Puranas the Asuras
were dictated to by the Planet Venus while the Devas were following the
Planet Jupiter.  Indeed the Indus valley calendar, as is well known was
dictated by the Jovian cycles\cite{r11}, a tradition which continues in
India even today. The well known Kumbha Mela begins with the entry of
Jupiter into Aquarius (Kumbha), and interestingly around 3000 B.C. and the
beginning of the traditional Indian calendar, this event took place when in
addition the Winter Solstice was also in Aquarius, a circumstance which
could have contributed to the beginning of the calendaric epoch of that time.\\
On the other hand it is equally remarkable that the Central American Mayan
civilization was indeed anti podal, that is they would be Asuras, and
moreover unlike any other known civilization they had a meticulous calendar
dictated by the Planet Venus, as is documented in the Dresden Codex\cite{r2}.\\
The Naga tribes alluded to above, have been associated with the Mayans and the netherworld or
the Asuras or shorn of allegory, the anti podal world. Indeed the surviving
Naga tribes of North Eastern India do share the mongloid features along
with the Mayans. Interestingly the names and symbols for numerals of the
Mayans and the ancient Naga tribes seem to be practically identical:
Hun (1), denoted by a $.$, ca (2), denoted by $..$, Ox (3) denoted by $...$, can/san
(4) denoted by $....$, ho (5) denoted by a horizontal bar and so on.
It is also reported\cite{r12} that the ancient Mayas chronicled an invader
from far off who was ambidextrous in his warrior skills, a description that
exactly describes the hero Arjuna of Mahabharata -one of his epithets is
Savya Sachi, or one who is equally adept with both hands. The significance
of the liason between Arjuna and the Nagas can now be seen.\\
Thus the above arguments make a case for a contact between Indus valley
India and the Mayan or a pre Mayan civilization (for example Olmec) of Central
America sometime around 1400 B.C. Incidentally this would also give the answer
to a long standing puzzle, namely evidence of the Indus script in the Easter
islands which are enroute\cite{r13}.

\end{document}